# Wettability of graphene by molten polymers

Maria Giovanna Pastore Carbone[1], Daniele Tammaro[3], Anastasios C. Manikas[1,2], George Paterakis[1,2], Ernesto Di Maio[3*] and Costas Galiotis[1,2*]

[1]Foundation of Research and Technology Hellas, Institute of Chemical Engineering and High Temperature Processes, Stadiou St. Platani, GR-26504 Patras (Greece)

[2]Department of Chemical Engineering, University of Patras, GR-26504 Patras (Greece)

[3]Dipartimento di Ingegneria Chimica, dei Materiali e della Produzione Industriale, University of Naples Federico II, P.le Tecchio 80, 80125 Naples (Italy)

**Abstract**

*Graphene wetting by polymers is a critical issue to both the success of polymer-aided transfer of large size sheets onto specific substrates and to the development of well performing nanocomposites. Here we show for the first time that high temperature contact angle measurements can be performed to investigate the wettability of CVD graphene by molten polymers. In particular, poly(methyl methacrylate), a widely used polymer support for CVD graphene transfer, has been adopted herein for this proof- of- concept study and the values of contact angle and work of adhesion have been provided in the temperature range 170-200 °C.*

*Corresponding authors:
Costas Galiotis, Tel.: +30-2610-965 255, fax: +30-2610-965 275, e-mail: c.galiotis@iceht.forth.gr

Ernesto Di Maio, Tel.: +39 081 768 25 11; fax: +39 081 768 24 04, e-mail: edimaio@unina.it

1. Introduction

Graphene, the perfect lattice of sp$^2$-bonded carbon atoms, cannot stand alone in real applications due to its atomic thickness and structure, and occurs either in combination with various substrates to form graphene-on-substrate devices or embedded in matrices to form nano-composites. In all these systems, the interface between graphene and the surrounding matrix or the underlying substrate plays a dominant role in both the manufacturing process and in effectively combining the properties of each component in the final structure. Regarding the production of graphene-polymer composites, for instance, it is well known that the control of wetting of molten polymers at solid surfaces and the adhesion forces involved at polymer-filler interfaces are key factors in the preparation of high-quality, high-performing composites [1]. As for the manufacturing of graphene-based devices, a transfer step is usually required to transfer the graphene sheet (either mechanically exfoliated or grown by Chemical Vapour Deposition, CVD) to an arbitrary substrate, and it mostly involves the use of a polymer that protects the one-atom-thick graphene sheet and allows for easy positioning on specific location of the substrate. In the transfer step, it has been inferred that the wetting and the adhesion of the protecting polymer layer to graphene may strongly influence the effectiveness and the quality of the deposition [2, 3]. In light of the above, it is obvious that investigating and understanding the wettability and the mechanism of adhesion between graphene and polymers can be crucial for both fundamental research and practical applications of graphene. In general, the interface between polymers and graphene has been largely investigated by using Raman spectroscopy [4]; moreover, several approaches based on the analysis of surface properties of CVD graphene and of polymers have been proposed to try to put light on the mechanism of adhesion [5, 6]. However, to the best of our knowledge, studies focused on the direct evaluation of the wettability of graphene by molten polymers have never been conducted.

In general, the wettability of graphene and its surface properties have been widely investigated by means of static contact angle measurements of low molecular weight liquids (e.g. water) and the "wetting transparency"– i.e. the transmission of the substrate wetting property over graphene coating – has gained significant attention due to its versatility for potential applications and has been a controversial issue in the literature [7]. Actually, more sophisticated studies have shown that graphene is intrinsically hydrophilic but its wettability may be dramatically influenced by the surrounding environment and by the substrate [8]. Recently, Raj et al. [9] have proposed the use of dynamic contact angle measurements combined with detailed graphene surface characterization and have demonstrated that the defects present in CVD grown and transferred graphene may result in unusually high contact angle hysteresis. In particular, this study highlighted for the first time that the only true representation of a graphene-coated surface is the advancing contact angle, while the receding contact angle is governed by defects and leads to significant contact angle hysteresis.

Contact angle (CA) testing is a generally simple characterization method: consider a liquid drop resting on a flat, horizontal solid surface, the contact angle is defined as the angle $\theta$ formed by the intersection of the liquid-solid interface and the liquid-vapour interface (geometrically acquired by applying a tangent line from the contact point along the liquid-vapour interface in the droplet profile) (Figure 1a). The interface where solid, liquid, and vapour co-exist is referred to as the "three-phases contact line". To determine the liquid surface tension from contact angle measurements, numerous methods have been developed. Ideally, the shape of a liquid drop depends on the combined effects of interfacial and gravitational forces. The balance between these forces is reflected mathematically in the Laplace equation of capillarity, which offers the possibility of determining surface tension by analysing the drop shape. Despite the apparent simplicity of sessile drop experiments, the experimental evaluation of reliable contact angle values at high temperatures remains a major problem and a serious obstacle to the development of scientific approaches to wetting phenomena [10]. High temperature measurements require a more complex experimental set-up since a small piece of solid sessile drop material is placed on a microscopically flat substrate and then heated above its melting temperature. Drops of low molecular weight liquids, e.g., water, ethyl alcohol, etc., rapidly achieve equilibrium shapes and can be readily observed and measured. In the case of polymer sessile drops on solid surfaces, the equilibrium contact angle is approached within a time interval that varies from minutes to hours, depending on the temperature and the system examined [10]. Therefore, in the investigation of wetting and spreading of polymer melts, it can be useful to distinguish between the contact angle measured during transient which is referred as "dynamic contact angle" (i.e. between the solid surface and the advancing liquid interface at the three-phase contact line) and the contact angle achieved at prolonged time of contact which is referred as "static contact angle" or "equilibrium contact angle" [11-14].

By using a versatile lab-scale mini batch [15], tailor-made in a two-view cell configuration and equipped with a camera for optical observation of the sessile drop, we have performed high temperature contact angle (HT-CA) measurements to investigate the wettability of CVD graphene by molten polymers (Figure 1b). In this configuration, the kinetics of wetting were followed by the rate of approach of the apparent dynamic contact to its final equilibrium value. Poly (methyl methacrylate) (PMMA), a widely used polymer support for CVD graphene transfer, has been adopted herein for this proof-of-concept study and measurements in the temperature range 170-200 °C were performed.

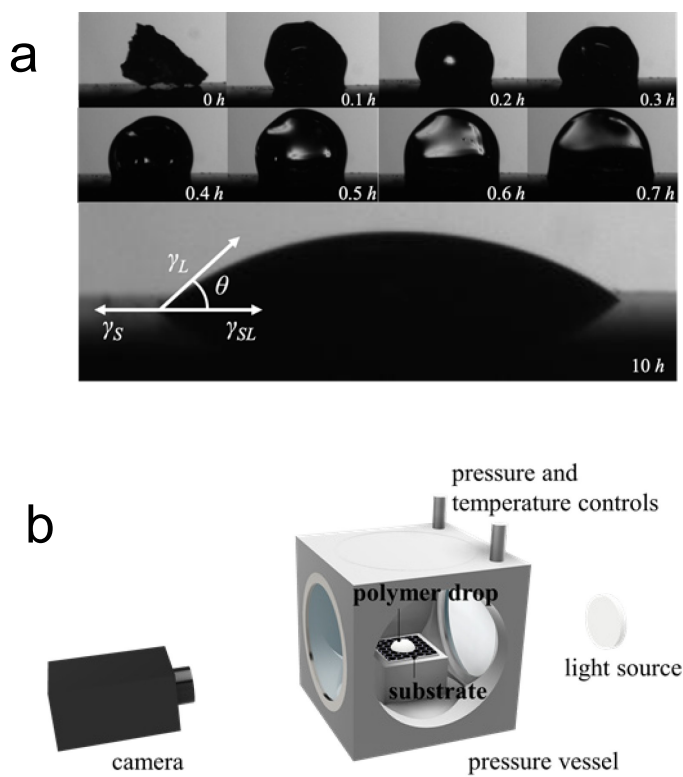

Figure 1 – Experimental setup: evolution of the polymer granule to form a sessile drop and the interfaces of a liquid droplet on a solid surface in the equilibrium configuration (a) and schematic of the apparatus adopted to measure contact angle of molten polymers on graphene (b).

## 2. Experimental

### 2.1 Materials

Graphene produced by CVD on 25-μm-thick Cu foil gently supplied by Aixtron SE (Herzogenrath, Germany) and a poly(methyl methacrylate) (PMMA, Sigma Aldrich, $M_W$ ~ 120000 by GPC, $M_w/M_n$ = 2.0-2.4) in granular form were adopted in this study. According to manufacturer, the composition of the PMMA is 6% isotactic, 39% atactic, and 55% syndiotactic and the $T_g$ measured by DSC is ~ 105° C (midpoint).

**2.2 Graphene transfer**

A thin film of PMMA (150nm ca.) was spin cast from a 3% wt solution in anisole (Sigma Aldrich) to form the protecting polymer layer on CVD graphene grown on Cu foil (typically 1 cm$^2$). The film was then annealed at 120 °C for 10 min to remove the solvent. Oxygen Plasma was used to remove graphene from the back side of the Cu foil. Cu was then etched away by using ammonium persulfate (APS) 0.15 M. After rinsing the film with distilled water, the target substrate was placed in water with a tilting angle underneath the floating film. Water was pulled out with a syringe to lower the film onto the substrate while positioning the film with a needle. After drying it under vacuum for several hours, the sample was heated at 180 °C in air for over 30 min to enable the flattening of the graphene film. The PMMA was finally removed with an acetone bath.

**2.3 Graphene characterization**

The success of the transfer and the quality of graphene transferred on Si wafer have been evaluated by using Raman mapping and Atomic Force Microscopy (AFM). Raman mapping has been performed on an area of 100x100 μm$^2$, by acquiring spectra at steps of 2μm (Renishaw Invia Raman Spectrometer, laser line 514 nm). Peaks were fitted using Lorentzian functions and the spectroscopic parameters (peak position and FWHM) were mapped. AFM was carried out with a commercial AFM (Dimension icon, Bruker Co., USA), in a Peak Force Tapping using silicon nitride tip (Scanasyst-Air, Bruker, nom. tip radius 2 nm, nom. frep. 70 kHz, nom. spring constant of 0.4 N/m) and with scan resolution of 384 samples per line. Height mode images are acquired simultaneously at a fixed scan rate (0.9 Hz) with a resolution of 384 × 384 pixels. Roughness of the CVD graphene samples transferred on the Si wafer was determined by AFM to investigate its effect on the observed contact angle by analysing images of size of 40 x 40 μm (i.e. projected surface area of 1600 μm$^2$). Processing of images and analysis was carried out using NanoScope™ software (Digital Instruments, version V614r1). All height images are presented after the first-order two-dimensional flattening.

## 2.4 Contact angle measurements

The sessile drop method was used to measure the contact angle of two couple PMMA/graphene and PMMA/silicon. All the measurements were performed at high temperature (from 170 to 200°C) and under vacuum to avoid oxidative degradation.

The setup designed for the experiment is shown in Figure 1. The polymer drop is laid on a horizontal substrate (i.e. graphene or silicon) in the middle of a closed metallic chamber that, in the present case, is a cross element model 15–24NFD cross from High Pressure Equipment Company, Erie, PA. This cross has four ports (based on ½'' NPT threads) with the following functions: two sight windows (for high temperature and pressure) in line with the drop to observe the contact angle, one venting port for sucking the air from the vacuum pump and one port dedicated to a platinum resistance thermometer (Pt100 with 3 wires) for measuring the temperature as close as possible to the polymer samples. The chamber is heated by two electrical squared resistance placed in contact with the upper and lower surface. The temperature control is given to a fuzzy thermo-regulator (Ascon-New England Temperature Solutions, Attleboro, MA, model X1) recorded via software on a computer.

The evolution process and the equilibrium contact angle were captured by a camera DMK 41AUO2 Germany, placed in front of the window 1 while the background is illuminated by a diffused light through the window 2 (as shown in Figure 1). The images were recorded and analysed by MATLAB software to evaluate the contact angle evolution as a function of time and temperature. The images were binarized using MATLAB function "im2bw" and the profile were traced using "bwboundaries" function. The contact angle was calculated by "telescope-goniometer" method [16] that is a direct measurement of the tangent line to the drop profile in the triple point.

For each experiment a spherical-shape piece of polymer of around 10 mg was gently laid in the centre of the substrate placed in the middle of the heated chamber. Care was taken to ensure that the substrate was level. Once the vacuum is reached, the temperature is increased to the maximum test temperature (i.e. 200°C) without overshoots and the evolution of the shape of the drop was continuously monitored by using the camera. Figure 1a reports a sequence of images of a characteristic experiment, showing the transient shapes towards the equilibrium, drop shape. After the equilibrium time, the shape of the drop is stable, and the contact angle does not evolve anymore and is recorded as the equilibrium value. At that time, the temperature is changed to study the temperature effect on the contact angle. As it will be observed in the following, equilibrium is reached after the characteristic time for the interfacial-tension-driven motions $t_c \approx \frac{\eta}{\gamma} R = 4 \cdot 10^4 s \approx 10h$ [17], where $\eta$ is the Newtonian viscosity of the PMMA at 200°C, $\gamma$ is the surface tension and $R$ is the sessile drop radius [18]. It is worth of note, furthermore, that, in our case, the ratio between the Laplace pressure and the hydrostatic pressure is much larger than 1 (the capillary length, $\sqrt{\frac{\gamma}{\rho g}} \gg R$, where $\rho$ is the fluid density) and the latter can be considered negligible [16, 19]. All the drops were visually checked after the test and no changes in colour were noticed.

## 3. Results and discussion

A large area (5x5 mm$^2$) of CVD monolayer graphene, as shown by the optical image in Figure 2a, was transferred from a copper foil to a Si/SiO$_2$ wafer by using the wet transfer technique. We could not locate any major cracks or fissures in the investigated samples. The quality of transferred graphene was verified by using Raman spectroscopy and AFM. An image from the latter (Figure 2b) shows typical features of CVD-grown graphene, namely, wrinkles and small amount of PMMA debris.

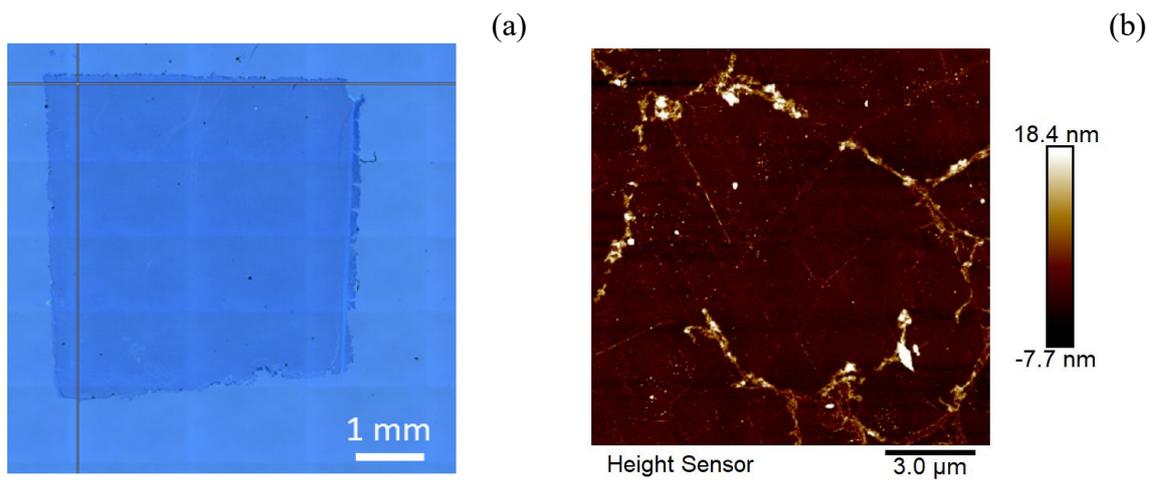

Figure 2 – Optical image (b) and AFM topography (b) of the monolayer CVD graphene transferred on $SiO_2$/Si wafer.

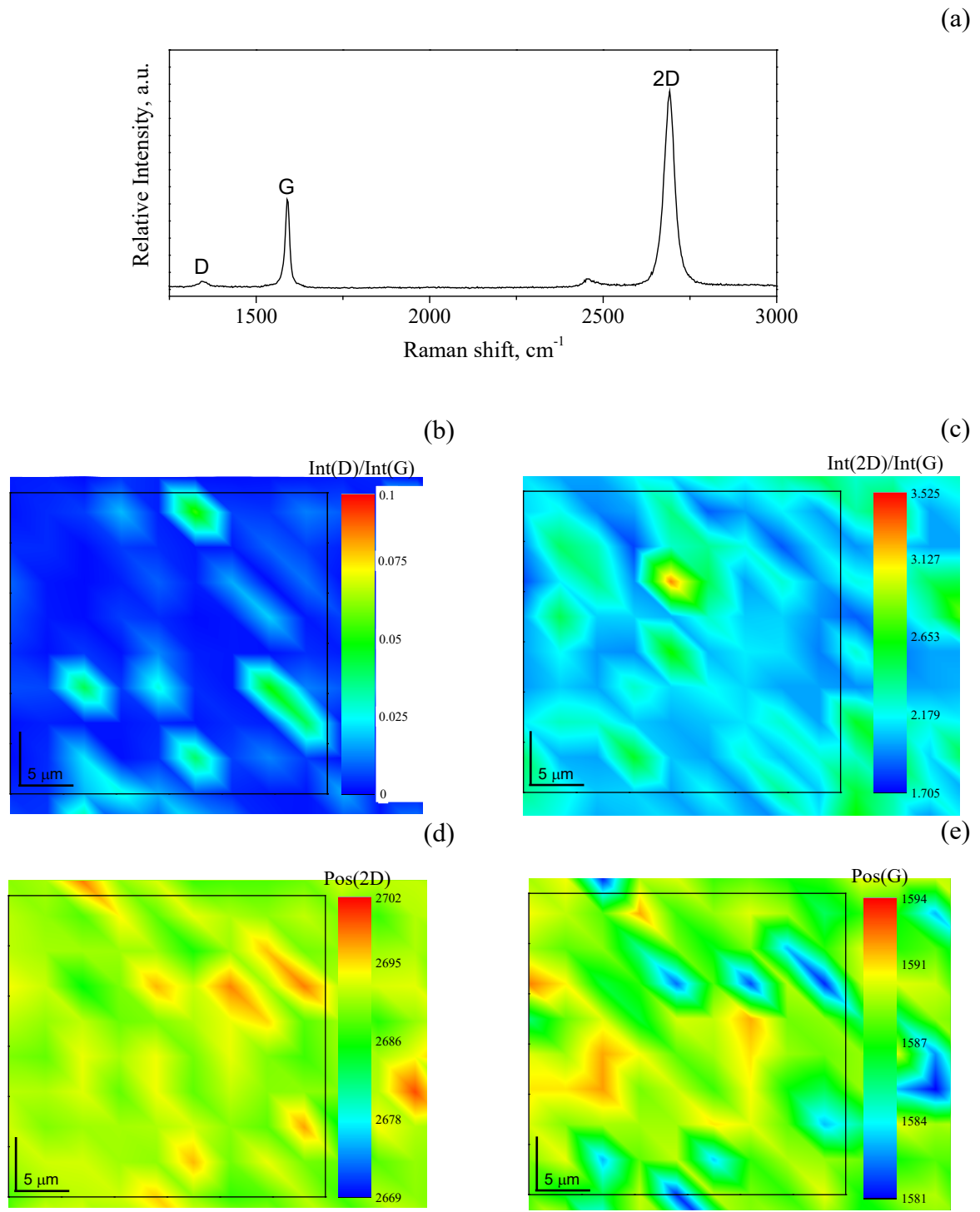

Figure 3 – Assessment of graphene quality after transfer on Si/SiO$_2$ wafer: representative Raman spectrum of CVD graphene after transfer (a) and spatial maps of I(D)/I(G) (b), I(2D)/I(G) (c), I(2D) (d), I(G) (e), Pos(2D) (f) and Pos(G) (g).

Representative Raman spectrum of transferred graphene is presented in Figure 3a, where the negligible D peak intensities and the large 2D/G peak ratio indicate that the as-synthesized CVD graphene is a high-quality single layer (Fig. 3b-c). As reported in Fig. 3d-e, the G and 2D band peaks of average values 1587.8±3 cm$^{-1}$ and 2691±3 cm$^{-1}$, respectively, appear slightly blue-shifted, as compared to perfect exfoliated graphene due to polymer shrinkage following the removal of solvent after spin coating.

The HT-CA experiments were carried out under vacuum to avoid thermo-oxidative degradation of the polymer. Once the vacuum was reached, the temperature was increased to the maximum test temperature (i.e. 200°C) and the evolution of the shape of the PMMA drop was continuously monitored by using the camera, as shown in Figure 1, until the attainment of a stable equilibrium shape of a spherical cap. The recorded images were analysed by MATLAB software to evaluate the evolution of the contact angle as a function of time. Measurements were repeated on 5 samples and at 170, 180, 190 and 200 °C. A typical temporal evolution of the cosine of the dynamic contact angle of molten PMMA on CVD graphene is shown in Figure 4a. As mentioned earlier, the wetting by polymer melts may require transient times which can be much higher than the observation time until the equilibrium drop shape is attained [10, 14]; during this transient, the contact area between the liquid drop and the solid may increase like in a typical advancing contact angle measurement. In particular, for the system at hand, the equilibrium value of contact angle is reached after several hours and that the duration of the transient is consistent with the characteristic time for the interfacial-tension-driven motions $t_c$ of 10h ca. (see section 2.4) Furthermore, the scattering of the long-time relaxation data depicted in Figure 4b reveals the pinning of the contact line on possible surface defects which lead to the "stick-slip" motion as "slips" or jumps with amplitude $\Delta sin\theta$. Comparing the temporal evolution of $sin\theta$ and the spreading velocity of the advancing sessile drop, a measure of the average distance between defect of 10μm can be estimated [17], which is consistent with the size of the graphene domains separated by typical wrinkles originated by copper grain boundaries, as shown in the AFM image (Figure 2b).

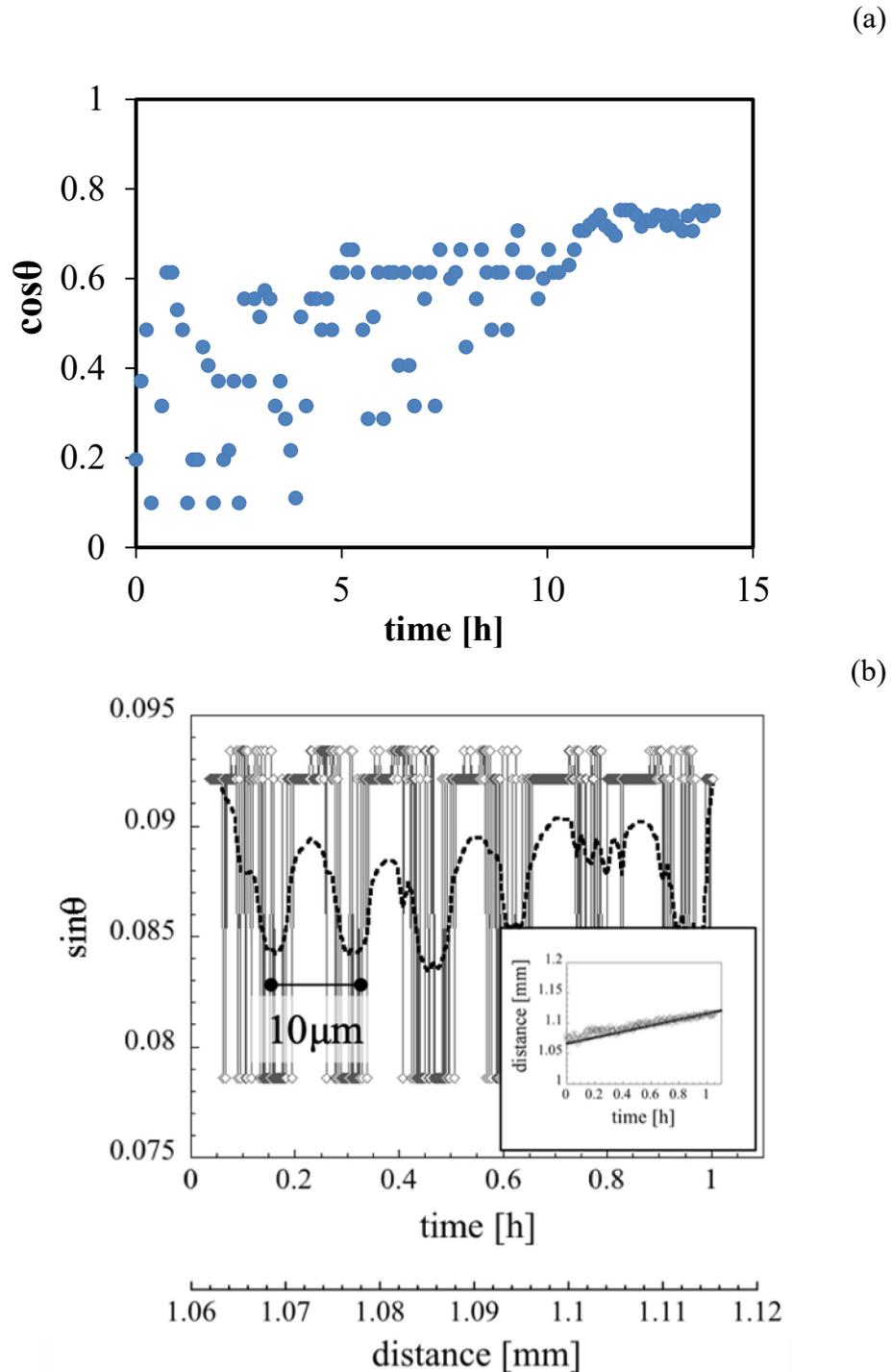

**Figure 4** Typical kinetics of wetting of PMMA on CVD graphene at 200° C: temporal evolution of the cosine of the dynamic contact angle θ (a) and long-time relaxation data (b).

The equilibrium values of the apparent contact angle of PMMA on CVD graphene as evaluated from the proposed method are reported in Fig.5a as a function of temperature. It is interesting noting that, in the investigated temperature range, it has been found to be almost constant, with a mean value of 31±2°, which is representative of a very high wettability. Furthermore, in order to assess any possible wetting transparency, HT-CA measurements were performed also on a PMMA drop located on a Si/SiO$_2$ wafer not covered with graphene. In this case, the contact angle of PMMA on silicon is about 41±2° in the investigated temperature range. It is important to note here that, on real surfaces, there may exist a wide range of metastable apparent contact angles [19] . However, as reported by Wolansky and Marmur [20], the most stable apparent contact angle is represented by the Wenzel contact angle, which takes into account the effect of the roughness of the investigated substrate. The Wenzel equation allows comparison of the experimentally determined apparent contact angle ($\theta_a$) and the Young's contact angle ($\theta_Y$) [21]:

$$\cos \theta_a = r \cos \theta_Y \quad \text{eq.1}$$

where $r$ is defined by the ratio of the apparent surface area to the projected surface area. The $r$ values found for our CVD graphene on Si/SiO$_2$ wafer was determined by AFM to be 1.000625. Using the mean value of $\theta_a$ of PMMA on graphene of 31°, $\theta_Y$ was determined to be 31.02°. This gives a difference between apparent and Young's CA less than 0.01%, which is within the uncertainty of the measurement and reveals that any error introduced by surface roughness is negligible in our system, as also demonstrated earlier for similar CVD graphenes [22].

The measured values of contact angle have been then used to calculate the work of adhesion between graphene and PMMA in the investigated temperature range. In general, it is known that the work of adhesion a solid surface and liquid – that refers to equilibrium energy associated with the reversible process of separation of the interface between two bulk phases - can be expressed as a function of the contact angle by

$$W_{SL} = \gamma_L (1 + \cos\theta) \qquad \text{eq.2}$$

where $\gamma_l$ represents the liquid surface free energy. The work of adhesion between graphene and PMMA evaluated by solving eq.2 is reported in Figure 5b in the investigated temperature range, and is found to slightly decrease from 0.055 to 0.051 N/m. In the solution of eq. 1, the values of surface tension of PMMA as reported by Wu [23] have been adopted for all the investigated temperatures. It is interesting noting that the values of work of adhesion for the system PMMA/CVD graphene estimated here on the bases of the proposed approach are in strict agreement with those estimated via AFM measurements for a similar system [24]. It is important to underline here that the thermodynamic work of adhesion would correlate with the strength of the adhesion; however, first of all, eq.2 is applicable to systems with secondary force interactions (no chemical bonds across the interface) and with no mechanical interlocking. Also, other important restriction is that, in the case of polymer-filler systems, the shrinkage stresses at the interface should be negligible and the polymer surface free energy should not change drastically after polymer solidification.

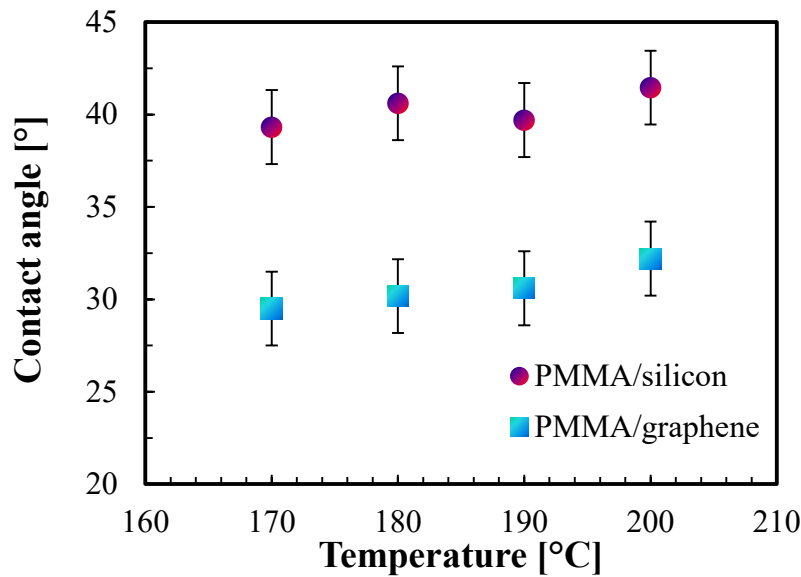

(a)

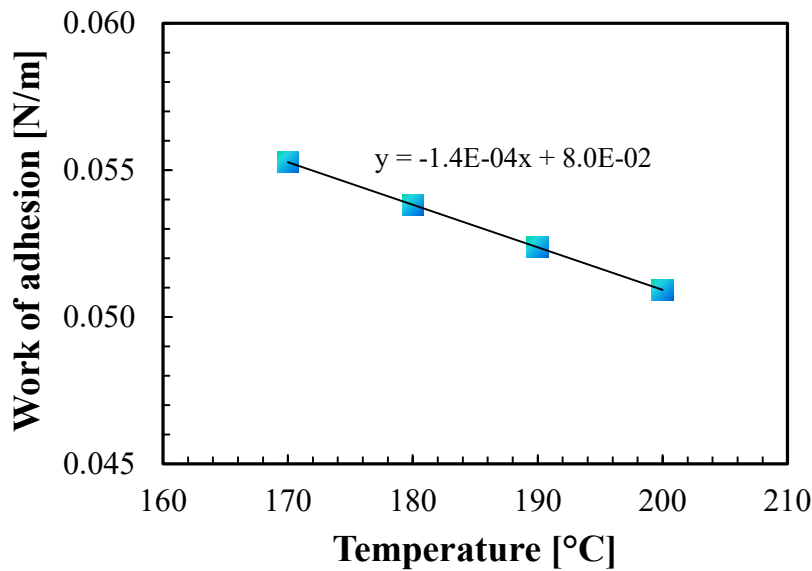

**Figure 5** Contact angle of molten PMMA on graphene and on SiO$_2$/Si wafer (a) and work of adhesion between graphene and PMMA (b) in the investigated temperature range.

## 4. Conclusions

In this study, for the first time the wetting behaviour of graphene by molten polymer has been investigated by performing high-temperature contact angle measurements. Wetting kinetics of molten PMMA drop on macroscopic CVD monolayer graphene were monitored and equilibrium values of apparent contact angle and work of adhesion have been provided. In the investigated temperature range, the contact angle was found to be almost constant, with a mean value of 31±2°, and the work of adhesion was found to slightly decrease, from 0.055 to 0.051 N/m. The technique can be easily exploited to investigate the wetting of different polymer melts on graphene and the results will provide valuable guidance for several applications, from the optimization of polymer-aided graphene transfer adopted in different fields (e.g. electronics) to the design and manufacturing of graphene-based composites.

**Acknowledgements**

The authors acknowledge the financial support of the European Research Council (ERC Advanced Grant 2013) via project no. 321124, "Tailor Graphene". The Open FET project "Development of continuous two-dimensional defect-free materials by liquid-metal catalytic routes" no. 736299-LMCat which is implemented under the EU-Horizon 2020 Research Executive Agency (REA) and is financially supported by EC and the General Secretariat for Research and Technology (GSRT) & the Hellenic Foundation for Research and Innovation (HFRI) are acknowledged for the financial support.